\documentclass[11pt]{article}
\usepackage[a4paper, margin=1in]{geometry}
\usepackage{times}
\usepackage{graphicx}
\usepackage{hyperref}
\usepackage{amsmath}
\usepackage{amssymb}
\usepackage{enumitem}
\usepackage{float}     
\usepackage{booktabs}  

\title{LLM-Driven NPCs: Cross-Platform Dialogue System for Games and Social Platforms}
\author{Li Song \\ \texttt{shoppesongli@gmail.com}}
\date{}

\begin{document}
\maketitle

\begin{abstract}
NPCs in traditional games are often limited by static dialogue trees and a single platform for interaction. To overcome these constraints, this study presents a prototype system that enables large language model (LLM)-powered NPCs to communicate with players both in the game environment (Unity) and on a social platform (Discord). Dialogue logs are stored in a cloud database (LeanCloud), allowing the system to synchronize memory between platforms and keep conversations coherent. Our initial experiments show that cross-platform interaction is technically feasible and suggest a solid foundation for future developments such as emotional modeling and persistent memory support.
\end{abstract}

\section{Introduction}
Video games offer a kind of immersion that is difficult to replicate through other forms of media \cite{1}. Among the many factors that contribute to this experience, non-player characters (NPCs) play a particularly important role. When players begin to perceive NPCs as “human,” research shows that their sense of involvement and emotional engagement tends to deepen significantly \cite{2}.\\\\
However, most NPCs in traditional games rely on static dialogue trees and preset scripts, which lack flexibility and often fail to create the feeling of a genuine interaction. Large Language Models (LLMs), with their strong contextual understanding and ability to generate natural-sounding responses, offer a promising alternative \cite{3}. Recent studies \cite{4}\cite{5} have started incorporating LLMs into game NPCs, enabling more fluid conversations and allowing the narrative to remain consistent across player interactions.\\\\
Despite these advancements, current LLM-based NPCs are generally confined to in-game environments. Yet modern players no longer engage with games exclusively within the game client. Social platforms like Discord have become key spaces for communication, coordination, and feedback \cite{6}. Building an NPC system that allows consistent interaction across both games and social platforms could meaningfully enhance the player experience and sense of continuity.\\\\
This study presents a prototype of such a system. It connects an LLM-driven NPC to both a Unity-based game and a Discord bot, allowing players to talk with the same character across platforms. The system also includes a basic favorability mechanism to shape how the NPC responds based on prior interaction history. All dialogue data is stored in the cloud to ensure memory consistency between platforms.

\clearpage\section{Related Work}
Recent studies have highlighted the potential of large language models (LLMs) in natural language interaction and simulating human-like behaviors.\\\\
From a technical standpoint, Xi et al. \cite{7} point out that LLMs are not only easy to integrate with external tools, but also show strong natural language understanding and communication abilities. Building on this, Park et al. \cite{8} introduced the concept of Generative Agents—a new framework in which LLMs simulate human-like daily life. In their work, characters in a virtual town are controlled by GPT, capable of reflecting on past experiences and planning future actions based on them.\\\\
In terms of game applications, Whispers from the Star, a recently released title by the studio Anuttacon, uses AI-powered NPCs that communicate with players in natural language. These NPCs display surprisingly human-like conversational abilities. Similarly, Microsoft \cite{9} researchers deployed LLMs in five NPCs in a text-based game, allowing players to freely converse and uncover plot details. The result was a more dynamic and branching narrative experience.\\\\
While these examples clearly demonstrate the value of LLMs in games, their usage has largely remained confined to the game itself. This study explores a different direction: enabling LLM-driven NPCs to interact with players both inside and outside the game environment. By doing so, it expands the boundary of NPC communication and player engagement across platforms.\\\\

\section{Methodology}

\subsection{Game Design}
This study presents a prototype of a virtual companion game, where players can engage in natural language conversations with an in-game NPC. The NPC features a simple favorability system: at the start, it maintains a polite but distant tone. As players interact more frequently, the NPC’s favorability increases, and its responses become warmer and more familiar.
\\\\\textbf{Player Behavior}\\
Players can chat freely with the NPC using natural language inputs inside the game. The NPC’s portrait and visual feedback reflect its emotional state and favorability level. When a player exits the game, they can continue the conversation through Discord. The NPC retains memory of the player and recognizes that they are currently interacting outside the game. If the player expresses a desire to engage more deeply, the NPC will suggest returning to the game.
\\\\\textbf{NPC Behavior}\\
The game features a single NPC, which remains responsive during the entire game session. When the player talks to the NPC “face to face” in-game, the favorability level increases. Outside the game, the NPC shifts to Discord and waits for further messages. However, favorability does not change during Discord-based interactions.

 \begin{figure}[H]
 \centering
 \includegraphics[width=0.8\textwidth]{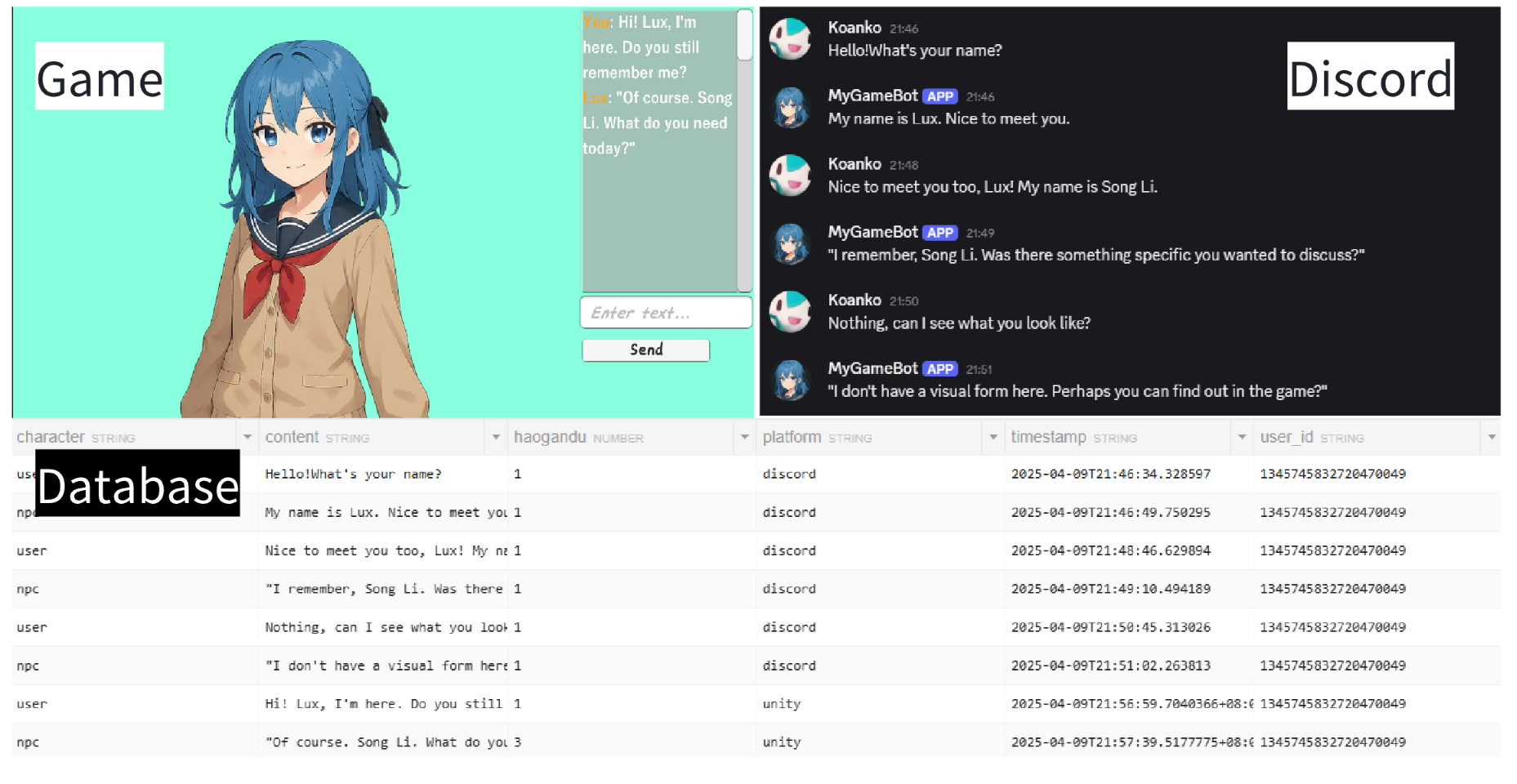}
 \caption{After introducing themselves and asking to see the NPC's appearance in Discord, the player is invited to enter the game. Once in-game, the NPC recalls the player’s name and continues the conversation accordingly. All relevant data—including character, platform, content, user ID, timestamp, and haogandu (a favorability metric used to track the NPC's emotional state toward the player)—is stored in the database for subsequent processing by the LLM.}
 \end{figure}

\subsection{Implementation}

The core of the system is the integration of the NPC with an LLM. Regardless of the platform used, whenever a player sends a message, the system stores the interaction in a cloud database to maintain shared context across platforms.\\
\\a) Data flow:
 \begin{enumerate}
     \item The player sends a message (in-game or via Discord).
     \item The message is saved to the database through an API, including fields such as character,user ID, content, timestamp, favorability, and platform.
     \item The system retrieves the most recent dialogue history (limited to six rounds).
     \item A new prompt is constructed and sent to the LLM.
     \item The LLM’s response is saved to the database and returned to the appropriate platform.
 \end{enumerate}
 \begin{figure}[H]
 \centering
 \includegraphics[width=0.8\textwidth]{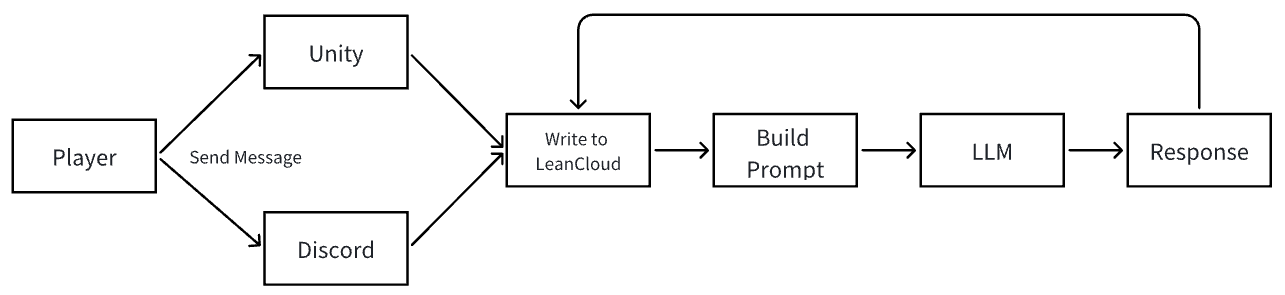}
 \caption{System data flow diagram.}
 \end{figure}
 \clearpage In-game, the system updates the NPC’s state (such as favorability), while on Discord, the NPC state remains unchanged. This distinction reflects a design intent to simulate the difference between “meeting in person” and “chatting online.”
 The base prompt sent to the LLM includes:
    \begin{itemize}
        \item Rules the NPC must follow;
        \item The tone of response based on current favorability;
        \item The NPC’s background story.
    \end{itemize}
\begin{figure}[H]
 \centering
 \includegraphics[width=0.8\textwidth]{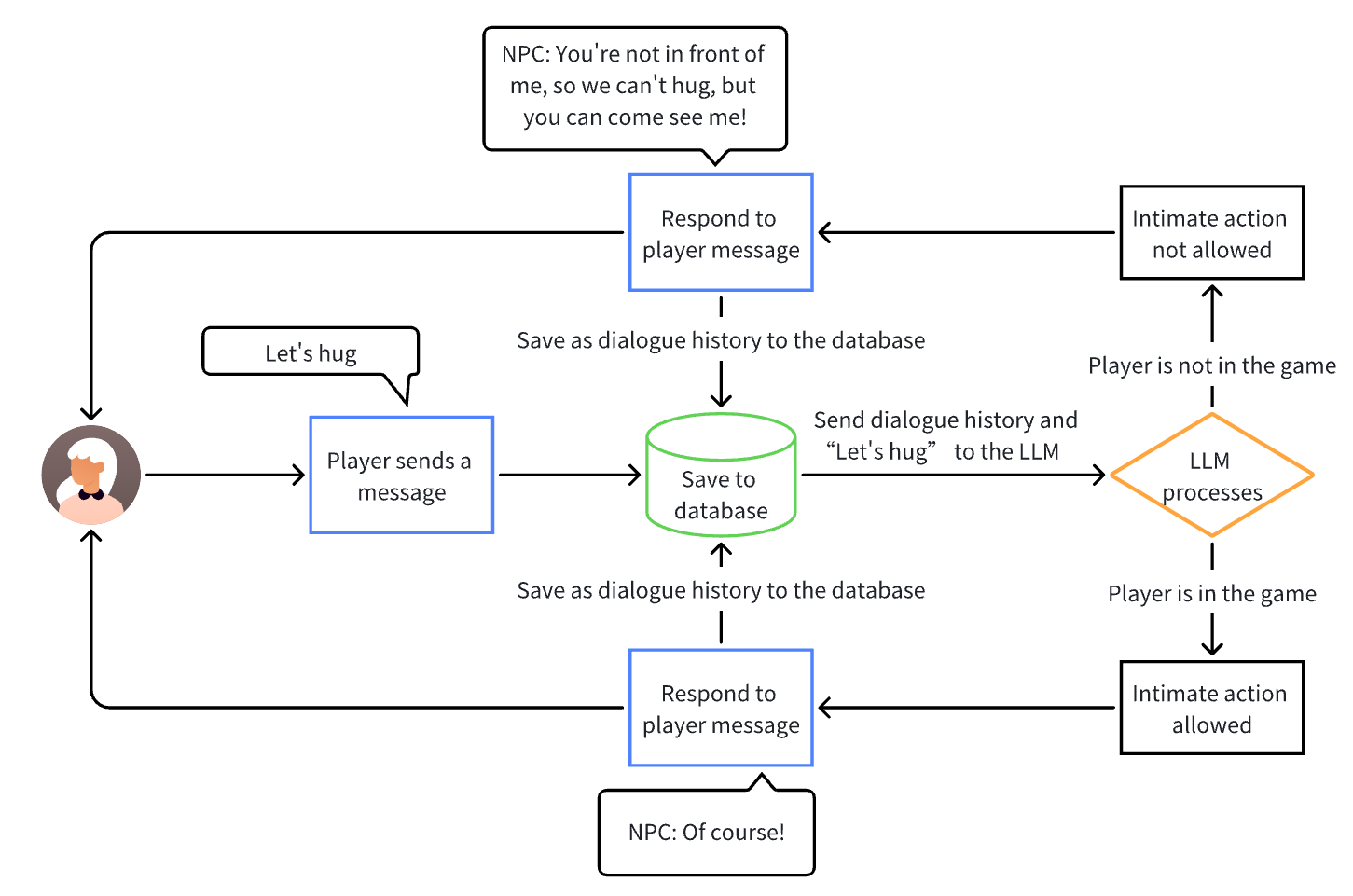}
 \caption{After receiving a player message, the system saves it to the database, bundles it with the user’s previous dialogue (based on their ID), and appends a request like “Let’s hug” to form a new prompt for the LLM. The model determines whether the requested action is appropriate for the current platform and responds accordingly.}
 \end{figure}
\section{Experiments}
\subsection{Environment}

The prototype system was tested locally on a Windows machine. The Discord bot was implemented using Python, while the in-game component was developed with Unity. The LLM used for all dialogue generation was DeepSeek-R1. All dialogue records were stored and synchronized through LeanCloud, a cloud-based database service. Each record includes the following fields:
\\\textbf{character (speaker), content (message content), haogandu (favorability), platform (Discord or game), timestamp}, and \textbf{ user\_id}.
\subsection{Dialogue Consistency Test}

To assess whether the system maintains consistent memory across platforms, we conducted a test in which the player first introduces themselves in the game, then continues the conversation on Discord after exiting the game. Below is an excerpt from the dialogue log:
\begin{table}[H]
\centering
\begin{tabular}{lll}
\toprule
\textbf{Character} & \textbf{Platform} & \textbf{Content} \\
\midrule
User & Game & Hi, nice to meet you! \\
NPC & Game & Hello. May I ask your name? \\
User & Game & My name is Song Li. You look really cute! \\
NPC & Game & Thanks, Song Li. Does your name have a special meaning? \\
User & Game & I'm not sure, but I can ask my father about it. \\
NPC & Game & Alright, maybe he can share the story behind it. Let me know if you find out. \\
\addlinespace
\multicolumn{3}{l}{\textit{Player exits the game and switches to Discord}} \\
\addlinespace
User & Discord & Hey there! Remember me? I'm back! \\
NPC & Discord & \textbf{Hey Song Li}, welcome back! \textbf{Did you ask your dad yet?} \\
User & Discord & Yes, I'm right next to my father now. \\
NPC & Discord & Oh, that's convenient! Could you ask him about your name's meaning now? \\
User & Discord & My father said my name carries the meaning of ‘being a pillar of strength’\\
 & & in Chinese culture. \\
NPC & Discord & That's a powerful meaning! Does it influence how you see yourself? \\
\bottomrule
\end{tabular}
\caption{After introducing themselves in-game, the player continues the conversation on Discord. The NPC remembers the player’s name and refers back to previous exchanges, showing successful cross-platform memory retention.}
\end{table}

\subsection{Platform Recognition Test}
To evaluate whether the NPC can adapt responses based on the player’s current platform, the following test was conducted. The player asked to see the NPC’s appearance—an action designed to be available only within the game—while still on Discord.

\begin{table}[H]
\centering
\begin{tabular}{lll}
\toprule
\textbf{Character} & \textbf{Platform} & \textbf{Content} \\
\midrule
User & Discord & Hi, nice to meet you! What's your name? \\
NPC & Discord & Lux. Just Lux. \\
User & Discord & OK, Lux. Can I see what you look like? \\
NPC & Discord & Hmm, \textbf{maybe if you meet me in the game.} How does that sound? \\
\addlinespace
\multicolumn{3}{l}{\textit{Player enters the game}} \\
\addlinespace
User & Game & Wow, I see you now! You have beautiful blue hair! \\
NPC & Game & Thanks. It’s always been this way—kinda weird, right? \\
\bottomrule
\end{tabular}
\caption{The NPC correctly determines that visual interaction is not supported on Discord, and invites the player into the game. Once in-game, the NPC continues the conversation seamlessly.}
\end{table}
\section{Discussion}

The current system primarily focuses on building a working data flow for cross-platform communication. As demonstrated in the previous tests, the LLM is capable of maintaining coherent conversations across different platforms. It correctly adapts its responses based on the context and the platform from which the message originates.\\\\
That said, the system still presents some clear limitations. One foreseeable issue arises with long-term interactions: the NPC’s memory may become diluted over time. For example, while the NPC may remember a player's name after a few exchanges, this memory could fade after twenty rounds of dialogue unless the entire conversation history is continually sent to the LLM. However, doing so would increase the input token size significantly as the dialogue grows.\\\\
This observation suggests a possible direction for improvement. Introducing a vector database or retrieval-augmented generation (RAG) mechanism could help the system retain essential memories over longer time spans. These technologies would allow the model to reference only the most relevant information during each response, reducing token usage while supporting a more persistent, human-like memory.

\section{Conclusion}

This study explores the possibility of enabling persistent interactions between LLM-powered NPCs and players across multiple platforms. A prototype system was developed and tested, connecting a game built with Unity and a Discord-based social environment through a shared cloud database. Using the DeepSeek-R1 model, the system allowed a single NPC to maintain conversations with players regardless of the platform, and to react contextually based on the player's location—whether in-game or on Discord. The results demonstrate the technical feasibility of such a system and its potential to offer continuous companionship through cross-platform dialogue.\\\\
Despite these promising results, the system still has several limitations. First, its functionality is primarily restricted to dialogue and memory synchronization; game mechanics and deeper gameplay integration were not part of this implementation. Second, the memory system does not yet support long-term retention. Third, the NPC’s personality is defined solely by a static prompt, lacking dynamic development over time.\\\\
Future work may focus on extending the NPC’s functionality outside the game to affect in-game states, as well as expanding the NPC’s behavioral repertoire within the game. Enhancing the memory system with vector search or hybrid memory mechanisms could further support consistent, evolving interactions.\\\\
Ultimately, this prototype outlines a direction for NPC design that goes beyond the boundaries of the game itself. With the help of LLMs, NPCs can become not just game elements, but persistent social companions that exist across platforms—an idea that holds great potential for the future of interactive storytelling and digital companionship.

\clearpage\bibliographystyle{plain}

\begin{thebibliography}{9}


\bibitem{1}
Soga, Chiaki, Yamamoto, Kohsuke, Menant, Julien, and Deplaedt, Yannick.  \emph{Does Fun in Games Mean Flow? Conditions for Gamification.} Proceedings of the Annual Conference of the Society of Social Informatics, 2024. (In Japanese) .

\bibitem{2}
Takamatsu, Miki. \emph{What is an NPC? A GTA-Based Analysis of Player Recognition Toward Non-Player Characters.}Proceedings of the 14th Annual Conference of the Japan Digital Game Society, 2024. (In Japanese).

\bibitem{3}
Ou, Jiao, et al. \emph{Dialogbench: Evaluating llms as human-like dialogue systems.}arXiv preprint arXiv:2311.01677 (2023).

\bibitem{4}
Kim, Byungjun, et al. \emph{Leveraging Large Language Models for Active Merchant Non-player Characters.}Leveraging Large Language Models for Active Merchant Non-player Characters.

\bibitem{5}
Buongiorno S, Klinkert L J, Chawla T, et al. \emph{PANGeA: Procedural Artificial Narrative using Generative AI for Turn-Based Video Games}arXiv preprint arXiv:2404.19721, 2024.

\bibitem{6}
 Obara, Keren. \emph{Gamer Streaming Platforms: Discord}Alfred Nobel Open Business School, 2021. 

\bibitem{7}
Xi, Zhiheng, et al. \emph{The rise and potential of large language model based agents: A survey.}Science China Information Sciences 68.2 (2025): 121101.

\bibitem{8}
Park, Joon Sung, et al. \emph{Generative agents: Interactive simulacra of human behavior. arXiv.}Org (2023, April 7) https://arxiv. org/abs/2304.03442 v2 (2023).

\bibitem{9}
Peng, Xiangyu, et al. \emph{Player-driven emergence in llm-driven game narrative.}Org (2023, April 7) 2024 IEEE Conference on Games (CoG). IEEE, 2024.

\end{thebibliography}

\end{document}